\def\rfr#1{eq. (\ref{#1})}

\def\cf#1#2{\dot\Omega^{\rm #2}_{.#1}}

\def\derp#1#2{\rp{\partial{#1}}{\partial{#2}}}
\def\bar{\begin{eqnarray}}
\def\ear{\end{eqnarray}}
\def\bb{\bibitem}
\def\eqi{\begin{equation}}
\def\eqf{\end{equation}}
\def\eqia{\begin{eqnarray}}
\def\eqfa{\end{eqnarray}}
\def\rp#1#2{{#1\over#2}}
\def\ct#1{\cite{#1}}
\def\lb#1{\label{#1}}

\def\oc2{$\mathcal{O}(c^{-2})$}

\documentclass[11pt]{article}
\usepackage{amsmath,amsthm,amscd,amssymb}
\usepackage{latexsym,rotating}
\usepackage{graphicx,epsfig}

\begin{document}

\noindent{\bf \LARGE{ Towards a $1\%$ measurement of the Lense-Thirring effect with LARES?  }}
\\
\\
\\
{Lorenzo Iorio}\\
{\it INFN-Sezione di Pisa, Italy\\}
Address for correspondence: {\it Viale Unit$\grave{a}$ di Italia 68, 70125, Bari, Italy
\\tel. 0039 328 6128815
\\e-mail: lorenzo.iorio@libero.it}

\begin{abstract}
After the recent approval by the Italian Space Agency (ASI) of the LARES mission, which will be launched at the end of 2008 by a VEGA rocket to measure the general relativistic gravitomagnetic Lense-Thirring effect by combining LARES data with those of  the existing LAGEOS and LAGEOS II satellites, it is of the utmost importance to  assess if the claimed accuracy  $\lesssim 1\%$ will be realistically obtainable.  A major source of systematic error is the mismodelling $\delta J_{\ell}$ in the static part of the even zonal harmonic coefficients $J_{\ell}, \ell=2,4,6,..$ of the multipolar expansion of the classical part of the terrestrial gravitational potential; such a bias crucially depends on the orbital configuration of LARES.  If for $\delta J_{\ell}$ the difference between the best estimates of different Earth's gravity solutions from the dedicated GRACE mission is conservatively taken instead of optimistically considering the statistical covariance sigmas of each model separately, as done so far in literature, it turns out that, since LARES will be likely launched in a low-orbit (semimajor axis $a\lesssim 7600$ km), the bias due to the geopotential may be up to ten times  larger than what claimed, according to a calculation up to degree $\ell=20$. Taking into account also the  even zonal harmonics with $\ell>20$, as required by the relatively low altitude of LARES, may further degrade the total accuracy. Should a nearly polar configuration (inclination to the Earth's equator $i\approx 90$ deg) be finally implemented, also other perturbations would come into play, further corrupting the measurement of the Lense-Thirring effect.
The orbital configuration of LARES may also have some consequences in terms of non-gravitational perturbations and measurement errors.

\end{abstract}



\section{Introduction}
The Italian Space Agency (ASI) recently made the following official announcement  \cite{ASIWEB}: ``On February 8, the ASI board approved funding for the LARES mission, that will be launched with VEGA's maiden flight before the end of 2008. LARES is a passive satellite with laser mirrors, and will be used to measure the Lense-Thirring effect.''  The italian version of the announcement yields some more information specifying that LARES, designed in collaboration with National Institute of Nuclear Physics (INFN), is currently under construction by Carlo Gavazzi Space SpA; its Principal Investigator (PI) is I. Ciufolini and its scientific goal is to measure at a $1\%$ level the general relativistic gravitomagnetic Lense-Thirring effect \cite{LT}, known also as frame-dragging, in the gravitational field of the Earth.

After what recently happened with the much more expensive Gravity Probe B (GP-B) \cite{Eve,GPB}, which should have measured another gravitomagnetic effect in the terrestrial gravitational field, i.e. the Schiff precession \cite{Schi} of four gyroscopes carried onboard at an accuracy much worse than the expected\footnote{See \cite{GPBWEB}.} $1\%$, the success of the cheaper LARES mission would be of great scientific significance.
%Unfortunately, the reality seems to be  different.

The major source of systematic uncertainty in such kind of satellite-based  measurements is represented by the impact of the static part of the even ($\ell=2,4,6...$) zonal ($m=0$) harmonic coefficients
$J_{\ell}$ of the multipolar expansion of the classical part of the terrestrial gravitational potential \cite{Kau} which induce noising effects having the same signature of the relativistic secular precessions of the satellite's node $\Omega$ and a much larger amplitude. Minimizing such a bias is of crucial importance and it strongly depends on the orbital parameters of LARES, especially the semimajor axis $a$ and the inclination $i$ to the Earth's equator. When it was proposed for the first time with the name of LAGEOS III \cite{Ciu86,Ciu89} the presently known LARES \cite{FaseA} had the same semimajor axis of the existing LAGEOS satellite ($a=12270$ km) and inclination supplementary to LAGEOS ($i_{\rm LAGEOS}=110$ deg, $i_{\rm LARES}=70$ deg); such a configuration would have allowed to use the sum of the nodes of both satellites enhancing the Lense-Thirring signal and cancelling out, at least in principle, the competing effect of all the even zonals.   The advent of recent models of the Earth's gravity field by the dedicated missions CHAMP \cite{CHAMPWEB}  and GRACE \cite{GRACEWEB} and the idea of combining LARES data with those of LAGEOS and LAGEOS II \cite{Ioretal02,IorNA05}, according to an idea put forth in \cite{Ciu96}, suggested to abandon  the original stringent requirements allowing for a lower orbit and a less narrow range for the inclination \cite{IorNA05}.

No details at all are released concerning the orbit in which LARES will be finally injected: ASI website says about VEGA that \cite{VEGAWEB}:
``[...] VEGA can place a 15.000 kg satellite on a low polar orbit, 700 km from the Earth. By lowering the orbit inclination it can launch heavier payloads, whereas diminishing the payload mass it can achieve higher orbits. [...]''    In the latest communication to INFN, Rome, 30 January 2008, \cite{INFN} I. Ciufolini writes that LARES will be launched with a semimajor axis of approximately 7600 km and an inclination between 60 and 80 deg.

As it will be shown, a detailed and realistic evaluation  of the impact of the bias due to the geopotential shows that such a low-orbit configuration should not allow to reduce such a systematic noise below the $1\%$ level being, instead, a few times larger.
%The crucial point resides in how the error due to the even zonals has been evaluated so far.

\section{An evaluation of the systematic bias due to the even zonal harmonics of the geopotential}
The Lense-Thirring effect consists of a tiny secular precession of the node $\Omega$ of  the orbit of a satellite moving around a central slowly rotating mass
\eqi\dot\Omega_{\rm LT}=\rp{2GS}{c^2 a^3(1-e^2)^{3/2}},\eqf
where $G$ is the Newtonian constant of gravitation, $S$ is the spin angular momentum of the central body, $a$ and $e$ are the semimajor axis and the eccentricity, respectively, of the satellite orbit. For the LAGEOS satellites it amounts to about 30 milliarcseconds per year (mas yr$^{-1}$).
 The observable to be used to measure it is the following linear combination of the nodes' rates \cite{IorNA05} $\Omega$ of LAGEOS, LAGEOS II and LARES \eqi \delta\dot\Omega^{\rm LAGEOS} + c_1\delta\dot\Omega^{\rm LAGEOS\ II}+c_2\delta\dot\Omega^{\rm LARES};\lb{combi}\eqf
$\delta\dot\Omega$ is an Observed-minus-Calculated (O-C) quantity for the satellites' nodal rates to be observationally constructed by processing the laser-ranging data without modelling the gravitomagnetic force.   It accounts, among other things, for all the unmodelled (like the Lense-Thirring effect itself) or mismodelled dynamical effects affecting the spacecraft nodes being equal to
$ X_{\rm LT}+\Delta,$
where $\Delta $ represents all the classical mismodelled/unmodelled
\ effects  of gravitational and non-gravitational origin.
The combined Lense-Thirring signature, which is expected to be present in the signal of \rfr{combi}, is
\eqi X_{\rm LT}=\dot\Omega^{\rm LAGEOS}_{\rm LT} + c_1\dot\Omega_{\rm LT} ^{\rm LAGEOS\ II}+c_2\dot\Omega_{\rm LT} ^{\rm LARES}.\lb{combilt}\eqf
The coefficients $c_1$ and $c_2$ entering \rfr{combi} and \rfr{combilt}     are defined as
\begin{equation}
\begin{array}{lll}
c_1 = \rp{\cf 2{LARES}\cf4{LAGEOS}-\cf 2{LAGEOS}\cf 4{LARES}}{\cf 2{LAGEOS\ II}\cf 4{LARES}-\cf 2{LARES}\cf 4{LAGEOS\ II}},\\\\
c_2 =  \rp{\cf 2{LAGEOS}\cf4{LAGEOS\ II}-\cf 2{LAGEOS\ II}\cf 4{LAGEOS}}{\cf 2{LAGEOS\ II}\cf 4{LARES}-\cf 2{LARES}\cf 4{LAGEOS\ II}}.
\end{array}\lb{cofi}
 \end{equation}
The quantities entering \rfr{cofi} are defined, for the satellite x, as
\eqi\cf\ell{x}=\derp{\dot\Omega_{\rm class}^{\rm x}}{J_{\ell}},\eqf
where \eqi\dot\Omega_{\rm class}=\sum_{\ell=2}\dot\Omega_{.\ell}J_{\ell}\eqf  is the classical node precession \cite{Kau} induced by the even ($\ell=2,4,6,...$) zonal ($m=0$) harmonic coefficients\footnote{They are defined as $J_{\ell}=-\sqrt{2\ell +1}\ \overline{C}_{\ell 0}$, where $\overline{C}_{\ell 0}$ are the normalized gravity coefficients.} $J_{\ell}$ of the multipolar expansion of the Newtonian part of the terrestrial gravitational potential; for example, for $\ell =2$ we have
\eqi\dot\Omega_{.2}=-\rp{3}{2}n\left(\rp{R}{a}\right)^2\rp{\cos i}{(1-e^2)^2},\eqf
where $R$ is the equatorial radius of the central body of mass $M$ and $n=\sqrt{GM/a^3}$ is the satellite's Keplerian mean motion.
The coefficients $\cf\ell{x}$, and thus also $c_1$ and $c_2$,    are functions of the semimajor axis $a$, the eccentricity $e$ and the inclination $i$ of the satellite x.   The combination of \rfr{combi}, along with \rfr{cofi}, is designed, by construction, to cancel out the mismodelled parts $\delta J_2$ and $\delta J_4$ of  the static and time-varying components of the first two even zonals, being affected by all the other ones of higher degree $\delta J_6, \delta J_8,...$

The evaluation of the systematic percent error $\delta\mu$ due to the geopotential has been  so far performed \cite{Ioretal02,IorGRG,IorNA05}
according to\footnote{Since we are dealing with a test of fundamental physics, one has  to be quite conservative. This is the reason why we do not consider a Root-Sum-Square evaluation of the systematic error and prefer to linearly adding the individual terms.}
\eqi\delta\mu\leq \left(\rp{\sum_{\ell=2}^{20}\left|\cf\ell{LAGEOS}+c_1\cf\ell{LAGEOS\ II}+c_2\cf\ell{LARES}\right|\delta J_{\ell}}{\dot\Omega^{\rm LAGEOS}_{\rm LT} + c_1\dot\Omega_{\rm LT} ^{\rm LAGEOS\ II}+c_2\dot\Omega_{\rm LT} ^{\rm LARES}}\right)100\lb{error}\eqf
by taking the covariance sigmas
%(more or less accurately calibrated)
 $\sigma_{\overline{ C }_{\ell 0}}$  of a given Earth's gravity solution and using them for the uncertainty $\delta J_{\ell}$ in the even zonals; in this way it has been always said that model $X$ yields a total error $x\%$,   model $Y$ yields a total error $y\%$ and so on. See, e.g., Table \ref{tavola1} for the EIGEN-GRACE02S solution.
\begin{table}
\caption{Systematic percent error $\delta\mu$ in the measurement of the Lense-Thirring effect with the combination of \rfr{combi} according to \rfr{error} and the calibrated covariance sigmas $\sigma_{\overline{ C }_{\ell 0}}$ of the Earth's gravity model EIGEN-GRACE02S up to degree $\ell=20$ for different values of the inclination $i$ of LARES ($a=7600$ km,  $e=0.001$). }
\centering
\begin{tabular}{|l|l|l|l|}
\hline
\multicolumn{1}{|c|}{} & \multicolumn{1}{c|}{$i=60$ deg} & \multicolumn{1}{c|}{$i=70$ deg} & \multicolumn{1}{c|}{$i=80$ deg}\\
\cline{1-4}
$\delta\mu$ (EIGEN-GRACE02S) & $4.3\%$ & $1.6\%$ & $3.8\%$\\
\hline

 \end{tabular}
\label{tavola1}

\end{table}
%
%
%
%
%
%It is precisely such an approach that induced people to thinking that combining LARES as in \rfr{combi} would allow to use a cheaper low-orbit %configuration.
Unfortunately, it is difficult to assess how realistically the sigmas of the covariance matrix of a given solution reflect the true uncertainties in the even zonals: in some cases, they are the mere formal, statistical errors, in other cases they are calibrated. However, also their calibration is often difficult to be realistically implemented, coming from a rather ad-hoc procedure. It is important to understand what kind of covariance information is taken for a
particular gravity field solution. A filter covariance, i.e.
the uncertainties are given as they are determined by the estimation filter, tends to yield rather optimistic errors. A more reliable approach$-$not followed so far in the publicly available models due to its great difficulties$-$would consist in using the so-called consider
covariance matrix, i.e. uncertainties of modelling parameters like, e.g., the
Earth's rotation, are taken into account. The consider covariance would give a rather
good indication of the real uncertainties.

\begin{figure}[htbp]
   \includegraphics[width=17cm,height=13cm]{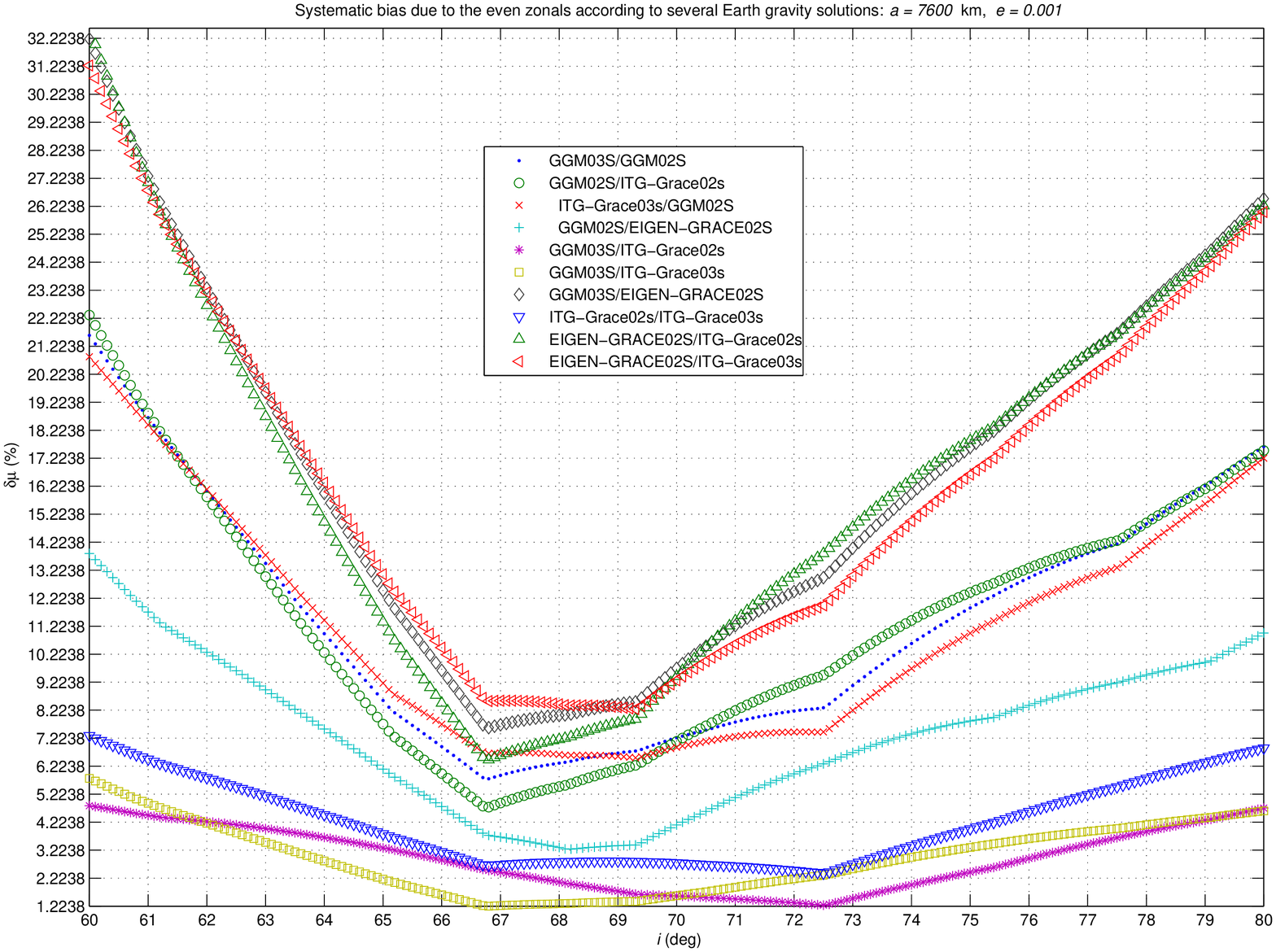}
   \caption{Systematic percent error $\delta\mu$ due to the even zonal harmonics up to degree $\ell=20$ for $a=7600$ km, $e=0.001$, $60$ deg $\leq i\leq 80$ deg according to several Earth's gravity models. The uncertainty $\delta J_{\ell}$ has been evaluated by taking the differences of the best estimates of the even zonals for several pairs of models. }
   \label{figura1}
   \end{figure}

\begin{figure}[htbp]
   \includegraphics[width=17cm,height=13cm]{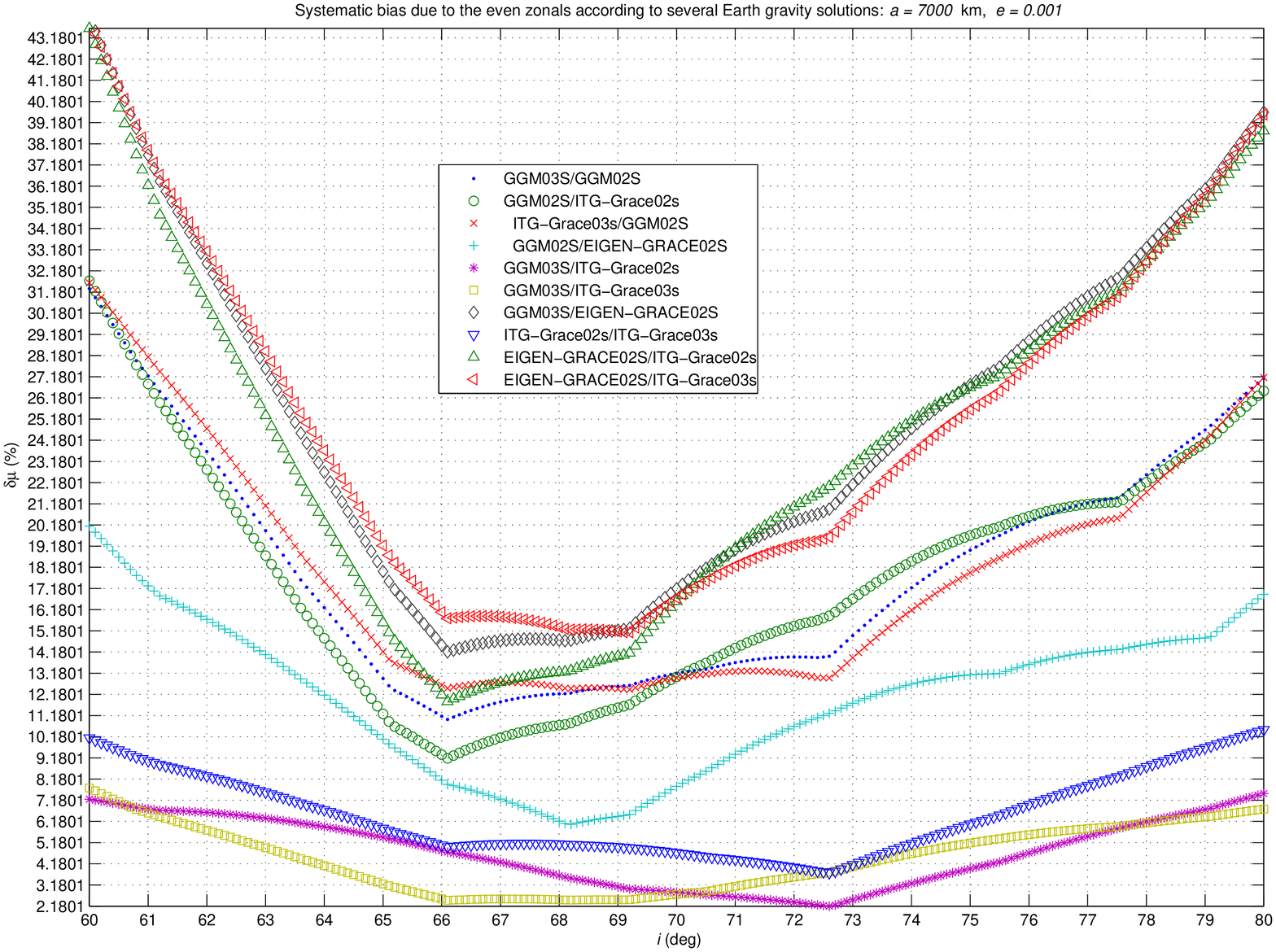}
   \caption{Systematic percent error $\delta\mu$ due to the even zonal harmonics up to degree $\ell=20$ for $a=7000$ km, $e=0.001$, $60$ deg $\leq i\leq 80$ deg according to several Earth's gravity models. The uncertainty $\delta J_{\ell}$ has been evaluated by taking the differences of the best estimates of the even zonals for various pairs of models. }
   \label{nuova}
   \end{figure}

A much more realistic and quantitative approach, which allows everyone to make an own idea, consists, instead, in taking for $\delta J_{\ell}$ the differences  $\Delta \overline{ C }_{\ell 0}=|\overline{ C }_{\ell 0}(X)-\overline{ C }_{\ell 0}(Y)|$ between the best estimates of the model $X$ and the model $Y$, which in many cases, are significantly larger than the sum of the sigmas $\sigma_{\overline{ C }_{\ell 0}}(X)+\sigma_{\overline{ C }_{\ell 0}}(Y)$.  Such an approach is usually applied by researchers involved in gravity field determination \cite{Lerch} and has been followed in the case of the previous tests of the Lense-Thirring effect with the LAGEOS satellites by Ciufolini himself \cite{Mandy,Ciu96}  by comparing the GEMT-3S and JGM-3 gravity fields. See also \cite{Lu07}.
In this paper we will follow it for several published models, retrievable at\footnote{Concerning the model GGM03S \cite{ggm03s}, I gratefully thank J. Ries, CSR, for having provided me with its spherical harmonic coefficients.} \cite{GFZ} along with references,  obtained by different institutions from GRACE data.
Concerning the choice of the models, we did not include in our analysis the latest solutions by GFZ encompassing data from LAGEOS  as well because of the strong possibility of {\it a-priori} `imprint' of the gravitomagnetic signature itself. Moreover, in view of the fact that the latest LAGEOS satellites SLR analyses concerning  the Lense-Thirring effect measurement (but not only in this case) have been based on satellite-only data, without considering the contribution of gravimetry and altimetry surface measurements, we did not considered solutions including such kinds of data like, e.g., EIGEN-CG03C.
It has been decided to use for LARES $a=7600$ km and $a = 7000$ km, and three different values for the inclination: $i=60,70,80$ deg. As can be noted, claiming a bias $\lesssim 1\%$ is unrealistic.
%If we take into account also the models based only on CHAMP data the situation is even less favorable, as shown by the last rows of Table %\ref{tavola2}.
The results are graphically depicted in Figure \ref{figura1} ($a=7600$ km) and Figure \ref{nuova} ($a= 7000$ km); see also Table \ref{rotata}.
%(in which no CHAMP-only models are present).

%
%
%
%
 %
 %
 %
 %
 %
 %
 %
 %
% For a lower orbit, i.e. $a=7000$ km, and $60$ deg $\leq i\leq 80$ deg the bias gets larger, as shown by Table \ref{rotata} and Figure \ref{nuova}.
%
%
%
 %
 %
 %
\begin{figure}[htbp]
   \includegraphics[width=17cm,height=13cm]{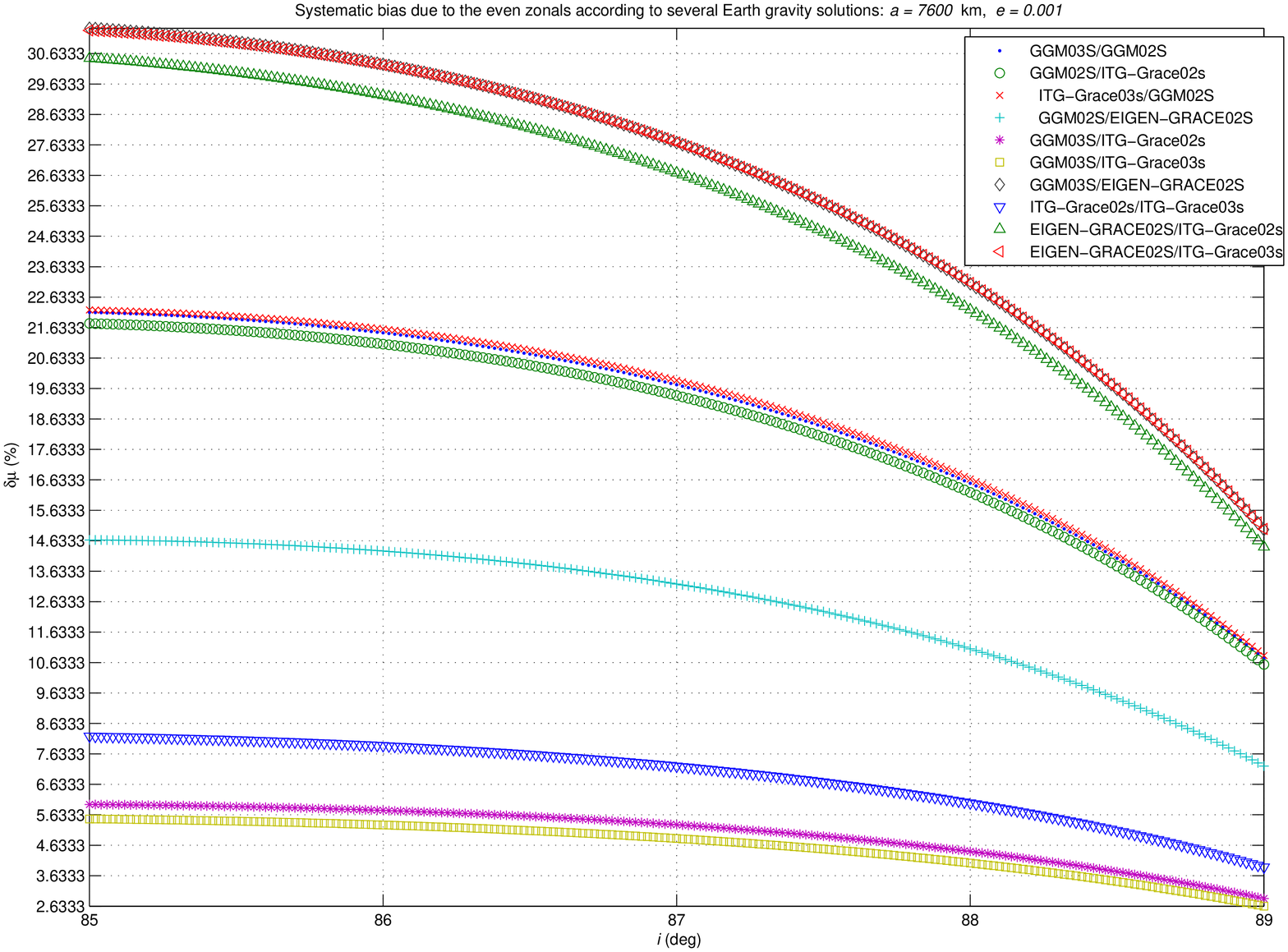}
   \caption{Systematic percent error $\delta\mu$ due to the even zonal harmonics up to degree $\ell=20$ for $a=7600$ km, $e=0.001$, $85$ deg $\leq i\leq 89$ deg according to several Earth's gravity models. The uncertainty $\delta J_{\ell}$ has been evaluated by taking the differences of the best estimates of the even zonals for various pairs of models.}
   \label{figura22}
   \end{figure}

\begin{figure}[htbp]
   \includegraphics[width=17cm,height=13cm]{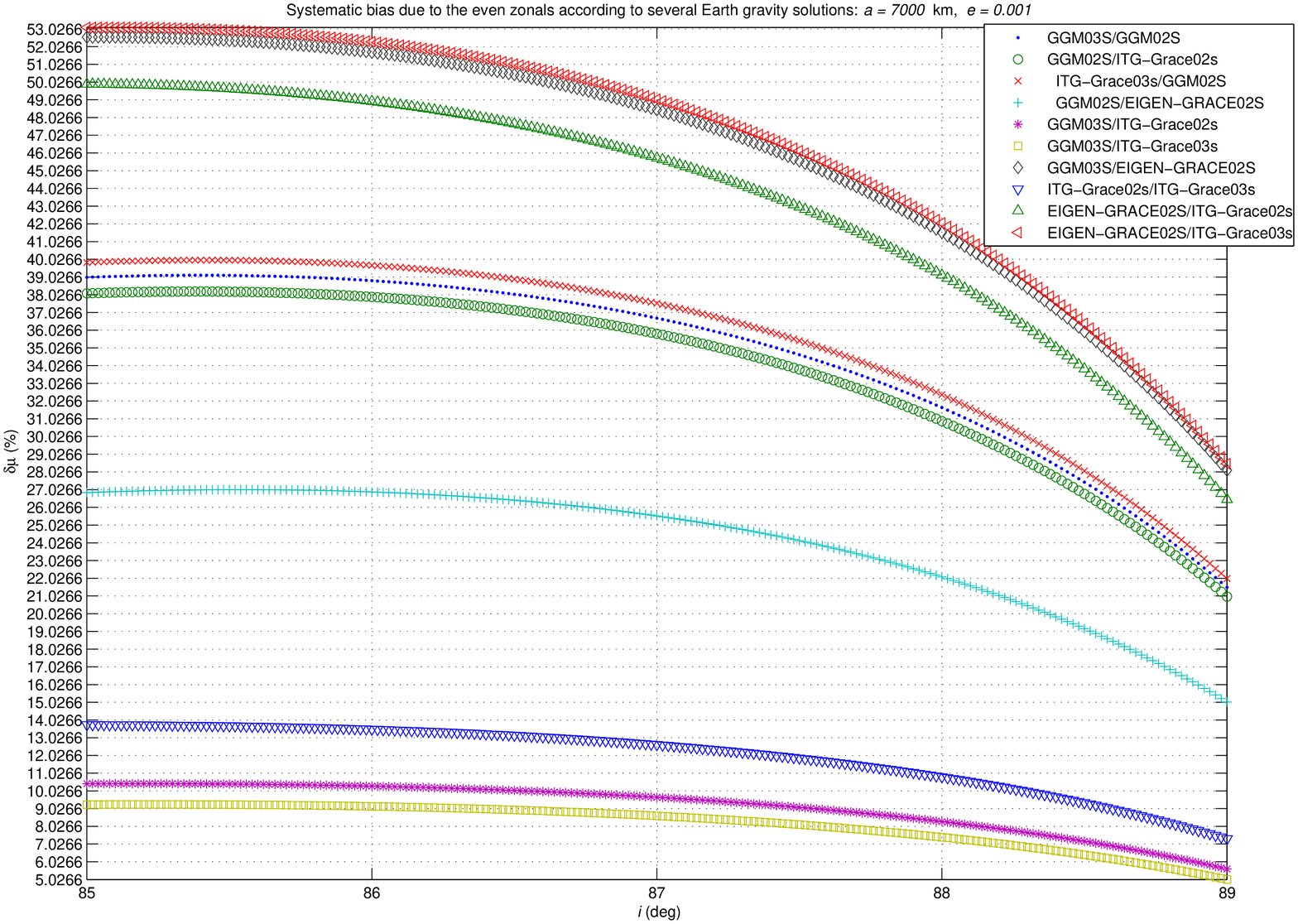}
   \caption{Systematic percent error $\delta\mu$ due to the even zonal harmonics up to degree $\ell=20$ for $a=7000$ km, $e=0.001$, $85$ deg $\leq i\leq 89$ deg according to several Earth's gravity models. The uncertainty $\delta J_{\ell}$ has been evaluated by taking the differences of the best estimates of the even zonals for various pairs of models.}
   \label{figura2}
   \end{figure}

\begin{sidewaystable}[h]
\caption{Systematic percent error $\delta\mu$ in the measurement of the Lense-Thirring effect with the combination of \rfr{combi} according to \rfr{error} and the difference $\Delta \overline{ C }_{\ell 0}$ among the best estimates for the even zonal coefficients for several Earth's gravity solutions up to degree $\ell=20$ for different values of the inclination $i$ of LARES and for $a=7600$ km, $a=7000$ km  ($e=0.001$). See also Figure \ref{figura1} and Figure \ref{nuova}. }
\centering
\begin{tabular}{|l|l|l|l|l|l|l|}
\hline
 &\multicolumn{2}{c|}{$i=60$ deg}&\multicolumn{2}{c|}{$i=70$ deg}&\multicolumn{2}{c|}{$i=80$ deg}\\
\cline{2-7}
 &$a=7600$ km& $a=7000$ km &$a=7600$ km& $a=7000$ km &$a=7600$ km& $a=7000$ km\\

\hline

$\delta\mu$ (EIGEN-GRACE02S-GGM02S)& $14\%$ & $20\%$ & $4\%$ & $8\%$ & $11\%$ & $17\%$\\

$\delta\mu$ (EIGEN-GRACE02S-GGM03S)& $32\%$ & $45\%$  & $10\%$ & $18\%$ & $26\%$ & $40\%$ \\

$\delta\mu$ (EIGEN-GRACE02S$-$ITG-Grace02s)& $33\%$ & $43\%$ & $9\%$ & $17\%$ & $26\%$ & $38\%$ \\

$\delta\mu$ (EIGEN-GRACE02S$-$ITG-Grace03s)& $31\%$ & $44\%$ & $9\%$ & $17\%$ & $26\%$ & $40\%$\\

$\delta\mu$ (GGM02S$-$GGM03S)& $22\%$ & $31\%$  & $7\%$  & $13\%$  & $18\%$  & $27\%$ \\

$\delta\mu$ (GGM02S$-$ITG-Grace02s)& $22\%$  & $32\%$  & $7\%$  & $13\%$  & $17\%$  & $26\%$ \\

$\delta\mu$ (GGM02S$-$ITG-Grace03s) & $21\%$  & $32\%$  & $7\%$  & $13\%$  & $17\%$  & $27\%$ \\

$\delta\mu$ (GGM03S$-$ITG-Grace02s)& $5\%$ & $7\%$ & $2\%$ & $3\%$ & $5\%$ & $7\%$ \\

$\delta\mu$ (GGM03S$-$ITG-Grace03s)& $6\%$ & $8\%$ & $2\%$ & $3\%$ & $5\%$ & $7\%$ \\

$\delta\mu$ (ITG-Grace03s$-$ITG-Grace02s)&   $7\%$ &   $10\%$ &   $3\%$ &   $5\%$ &   $7\%$ &   $10\%$\\

 \hline
\end{tabular}
\label{rotata}
\end{sidewaystable}
Analogous analyses conducted for $a=7600$ km and $a=7000$ and $i=85$ deg, $i=89$ deg, $i=91$ deg, which corresponds to the most likely orbital
configuration allowed by VEGA, show a neat worsening, as shown by Table \ref{rotata2}.
\begin{sidewaystable}[h]
\caption{Systematic percent error $\delta\mu$ in the measurement of the Lense-Thirring effect with the combination of \rfr{combi} according to \rfr{error} and the difference $\Delta \overline{ C }_{\ell 0}$ among the best estimates for the even zonal coefficients for several Earth's gravity solutions up to degree $\ell=20$ for different values of the inclination $i$ of LARES close to polar geometry and for $a=7600$ km, $a=7000$ km  ($e=0.001$). See also Figure \ref{figura22} and Figure \ref{figura2}. }
\centering
\begin{tabular}{|l|l|l|l|l|l|l|}
\hline
 &\multicolumn{2}{c|}{$i=85$ deg}&\multicolumn{2}{c|}{$i=89$ deg}&\multicolumn{2}{c|}{$i=91$ deg}\\
\cline{2-7}
 &$a=7600$ km& $a=7000$ km &$a=7600$ km& $a=7000$ km &$a=7600$ km& $a=7000$ km\\

\hline

$\delta\mu$ (EIGEN-GRACE02S-GGM02S)&         $14\%$ & $27\%$ &   $7\%$ &      $15\%$ &    $16\%$ &    $44\%$\\

$\delta\mu$ (EIGEN-GRACE02S-GGM03S)&         $32\%$ & $52\%$  &  $16\%$ &     $28\%$ &    $34\%$ &    $82\%$ \\

$\delta\mu$ (EIGEN-GRACE02S$-$ITG-Grace02s)& $30\%$ & $50\%$ &   $15\%$ &     $28\%$ &    $32\%$ &    $77\%$ \\

$\delta\mu$ (EIGEN-GRACE02S$-$ITG-Grace03s)& $31\%$ & $53\%$ &   $16\%$ &     $30\%$ &    $34\%$ &    $83\%$\\

$\delta\mu$ (GGM02S$-$GGM03S)&               $22\%$ & $39\%$  &  $11\%$  &    $21\%$  &   $24\%$  &   $63\%$ \\

$\delta\mu$ (GGM02S$-$ITG-Grace02s)&         $21\%$ & $38\%$  &  $10\%$  &    $21\%$  &   $23\%$  &   $61\%$ \\

$\delta\mu$ (GGM02S$-$ITG-Grace03s) &        $22\%$ & $40\%$  &  $11\%$  &    $22\%$  &   $24\%$  &   $64\%$ \\

$\delta\mu$ (GGM03S$-$ITG-Grace02s)&         $6\%$  & $10\%$ &   $3\%$ &      $5\%$ &     $6\%$ &     $16\%$ \\

$\delta\mu$ (GGM03S$-$ITG-Grace03s)&         $5\%$  & $9\%$ &    $3\%$ &      $5\%$ &     $6\%$ &     $15\%$ \\

$\delta\mu$ (ITG-Grace03s$-$ITG-Grace02s)&   $8\%$  & $14\%$ &   $4\%$ &      $7\%$ &     $9\%$ &     $21\%$\\

 \hline
\end{tabular}
\label{rotata2}
\end{sidewaystable}
 In Figure \ref{figura22} and \ref{figura2} we depict the situation for  $a=7600-7000$ km, $e=0.001$, $85$ deg $\leq i\leq 89$ deg.

 %
 %
 %
 %
 %
%It must be noted that an inclination too close to $i=90$ deg would be fatal because the coefficient $c_2$ of LARES in the combination of \rfr{combi}
%would become much larger than 1 (see Figure \ref{figura3}) greatly enhancing all time-varying perturbations of gravitational and non-gravitational %origin.
%
%
%
%

A caveat concerning the results presented so far is that, in principle, they might be   optimistic because they have been obtained by only accounting for the even zonals up to degree $\ell=20$. But, with such a small semimajor axis, it maybe required to take into account also the impact of the other even zonals of higher degree $\ell\gg 20$. Such doubts are, in fact, enforced by a straightforward calculation according to the standard approach by \cite{Kau} for $a=7600$, km $i=71$ deg, $e=0.001$ whose results are presented in Table \ref{tavolazza}.
\begin{table}
\caption{ Systematic percent error $\delta\mu$  in the measurement of the Lense-Thirring effect with LAGEOS, LAGEOS II and LARES according to \rfr{combi} and $\delta J_{\ell}= \Delta J_{\ell}$ up to degree $\ell = 60$ for the global Earth's gravity solutions  considered here for $a=7600$, km $i=71$ deg, $e=0.001$.}\label{tavolazza}
\centering
\begin{tabular}{|l|l|}
\hline %
\multicolumn{1}{|c|}{Models compared} & \multicolumn{1}{c|}{$\delta\mu(\%)$}\\
\cline{1-2}
\hline

EIGEN-GRACE02S$-$GGM02S & $36\%$\\
EIGEN-GRACE02S$-$GGM03S & $52\%$\\
%
%EGM2008$-$JEM01-RL03B & 8\%\\
%
%EGM2008$-$GGM02S & 27\%\\
%
%EGM2008$-$GGM03S & 5\%\\
%
%EGM2008$-$ITG-Grace02 & 4\%\\
%
%EGM2008$-$ITG-Grace03 & 0.1\%\\
%
%EGM2008$-$EIGEN-CG03C & 38\%\\
%
%EGM2008$-$EIGEN-GRACE02S & 53\%\\
%
%EGM2008$-$EIGEN-GL04C & 35\%\\
%
%JEM01-RL03B$-$GGM02S & $28\%$\\
%
%JEM01-RL03B$-$GGM03S & $10\%$\\
%
%JEM01-RL03B$-$ITG-Grace02 & $8\%$\\
%
%JEM01-RL03B$-$ITG-Grace03s & $8\%$\\
%
%JEM01-RL03B$-$EIGEN-CG03C & $44\%$\\
%
%JEM01-RL03B$-$EIGEN-GRACE02S & $57\%$\\
%
%JEM01-RL03B$-$EIGEN-GL04C & $41\%$\\
EIGEN-GRACE02S$-$ITG-Grace02s & $54\%$\\
EIGEN-GRACE02S$-$ITG-Grace03s & $53\%$\\
GGM02S$-$GGM03S & $24\%$\\
GGM02S$-$ITG-Grace02s& $28\%$\\
GGM02S$-$ITG-Grace03s& $26\%$\\
%
%GGM02S$-$EIGEN-CG03C & $27\%$\\
%
%GGM02S$-$EIGEN-GL04C & $24\%$\\
%
GGM03S$-$ITG-Grace02s & $5\%$\\
GGM03S$-$ITG-Grace03s & $5\%$\\
%
%GGM03S$-$EIGEN-CG03C & $36\%$\\
%
%GGM03S$-$EIGEN-GL04C & $33\%$\\
%
ITG-Grace02s$-$ITG-Grace03s & $4\%$\\
%
%ITG-Grace02$-$EIGEN-CG03C & $39\%$\\
%
%ITG-Grace02$-$EIGEN-GL04C & $37\%$\\
%
%ITG-Grace03s$-$EIGEN-CG03C & $38\%$\\
%
%ITG-Grace03s$-$EIGEN-GL04C & $35\%$\\
%
%EIGEN-CG03C$-$EIGEN-GRACE02S & $27\%$\\
%
%EIGEN-CG03C$-$EIGEN-GL04C & $8\%$\\
%
%EIGEN-GRACE02S$-$EIGEN-GL04C & $32\%$\\
%
\hline

\end{tabular}

\end{table}
%
%
%
%

%A reliable evaluation of such a potentially non-negligible source of systematic error is not an easy task because it turns out that, %even for Stella ($a=7193$ km) and Starlette ($a=7331$ km), the computation of the coefficients $\dot\Omega_{.\ell}$, performed with two different %softwares according to the scheme of \cite{Kau}, tend to wildly oscillate finally diverging for $\ell\approx 50$.  The impact of the higher-degree %even zonal harmonics on combinations involving also the other existing geodetic satellites was the subject of numerous papers   like, e.g., %\ct{multizon}.

\begin{figure}[htbp]
   \includegraphics[width=17cm,height=13cm]{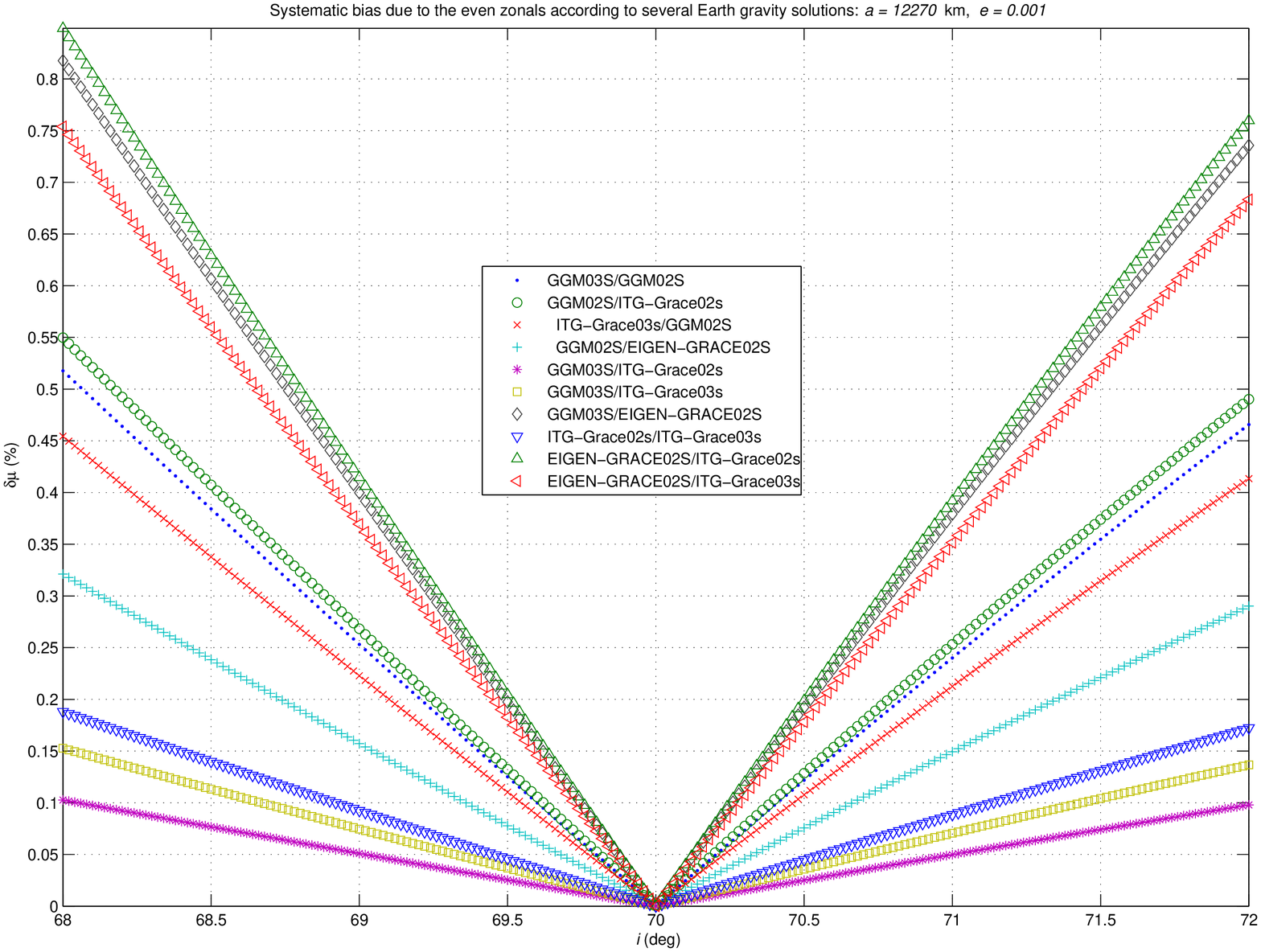}
   \caption{Systematic percent error $\delta\mu$ due to the even zonal harmonics up to degree $\ell=20$ for $a=12270$ km, $e=0.001$, $68$ deg $\leq i\leq 72$ deg according to several Earth's gravity models. The uncertainty $\delta J_{\ell}$ has been evaluated by taking the differences of the best estimates of the even zonals for various pairs of models.}
   \label{figura3}
   \end{figure}

The only valid solution would be to use the originally proposed orbital configuration for LARES ($a=12270$ km, $i=70$ deg), as shown by  Figure \ref{figura3}; as can be noted, for a deviation of 1 deg from the optimal choice $i=70$ deg the systematic error would be well below $1\%$.
\section{The coefficients of the combination}
The inclination in which LARES will be finally inserted into its orbit is also crucial in determining the size of the coefficient $c_2$ with which its node enters the combination of \rfr{combi}. It does matter because it may enhance all the perturbations not cancelled out by the coefficients of \rfr{cofi}. They are the non-gravitational ones, whose impact on the node of LARES should be reduced to $\approx 0.1\%$ of the Lense-Thirring effect thanks to its careful construction, and the $\ell=2,m=1$ component of the solar $K_1$ tide whose node harmonic perturbation has the same period of the satellite's node \ct{tide} and for which, of course, the physical properties of LARES are of no concern. It turns out that the case for 60 deg $\leq i\leq 80$ deg does not pose problems
 %
% \begin{figure}[htbp]
%   \includegraphics[width=17cm,height=13cm]{coef1.jpg}
%   \caption{Coefficients $c_1$ and $c_2$ of LAGEOS II and LARES for $a=7600$ km, $e=0.01$ and \rfr{cofi} for 60 deg $\leq i\leq 80$ deg. }
%   \label{coef1}
%   \end{figure}
 %
 %
 %
 %
because $c_1< 0.5$ and $c_2$ is of the order of $10^{-2}$.

\begin{figure}[htbp]
   \includegraphics[width=17cm,height=13cm]{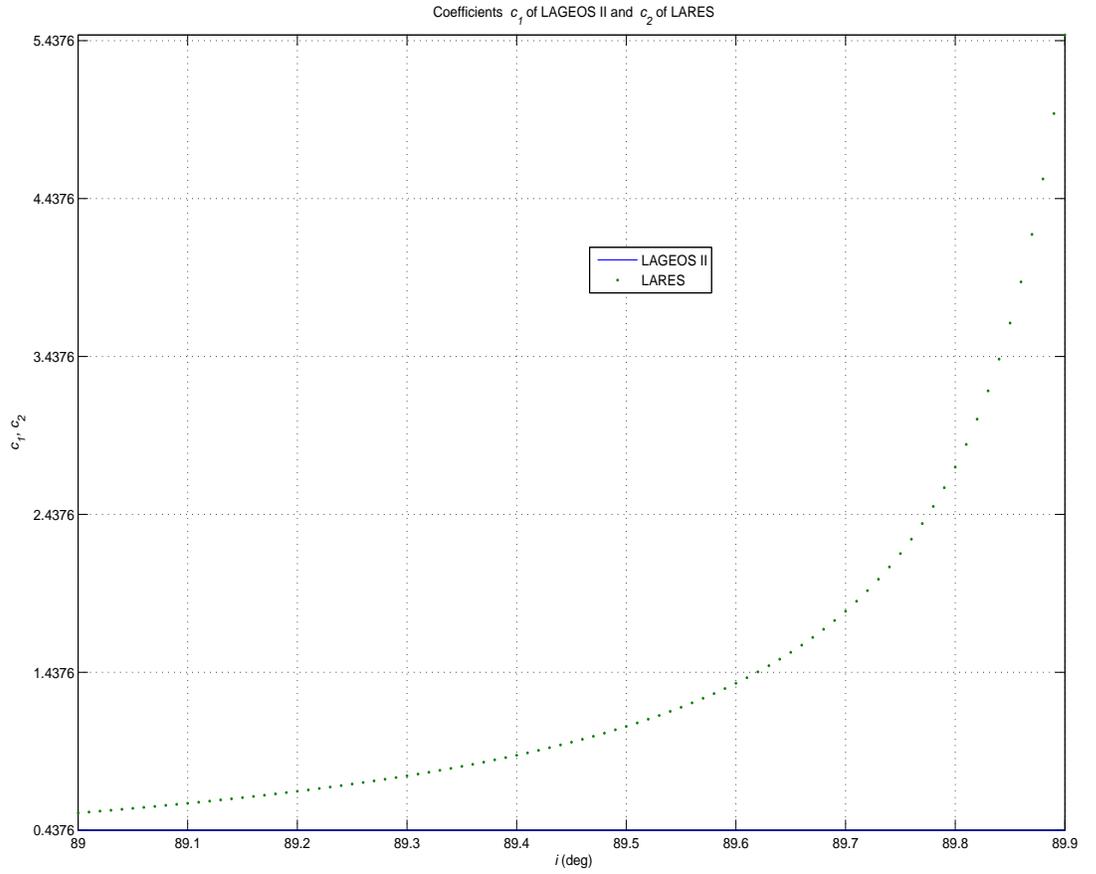}
   \caption{Coefficients $c_1$ and $c_2$ of LAGEOS II and LARES, according to \rfr{cofi}, for $a=7000$ km, $e=0.001$  and 89.0 deg $\leq i\leq 89.9$ deg. For 89.90 deg $\leq i\leq 89.99$ deg $c_2$ gets as large as 55.}
   \label{coef2}
   \end{figure}

\begin{figure}[htbp]
   \includegraphics[width=17cm,height=13cm]{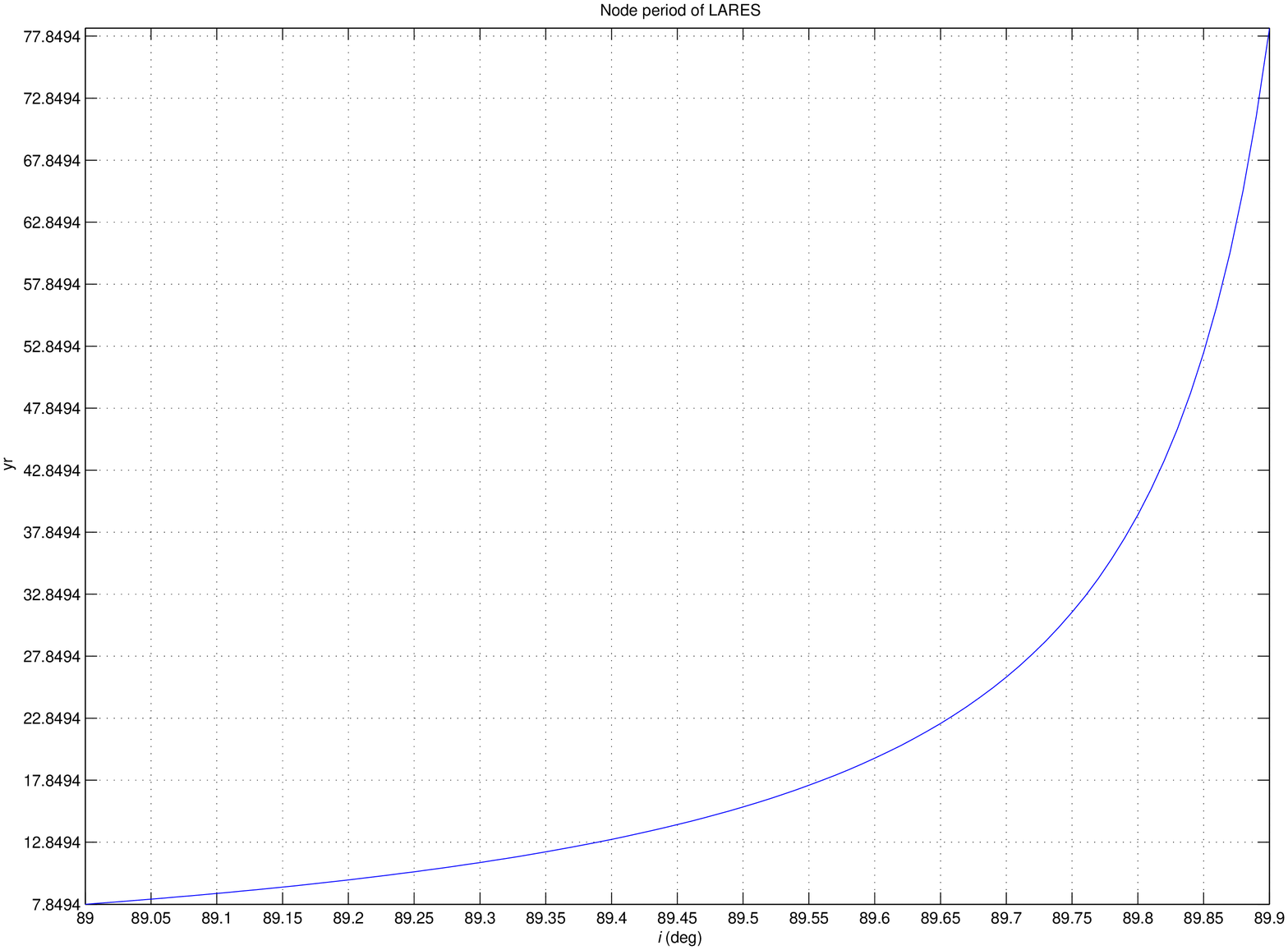}
   \caption{Period of the node of LARES, in years, for $a=7000$ km, $e=0.001$  and 89.0 deg $\leq i\leq 89.9$ deg. }
   \label{nodo}
   \end{figure}

The situation is quite different for inclinations close to $i=90$ deg for which $c_2$ diverges.
Figure \ref{coef2} shows   that for  $a=7000$ km and 89.0 deg $\leq i\leq 89.9$ deg, i.e. the most likely orbital configuration in which VEGA will deploy LARES, the coefficient $c_2$ of the node of LARES becomes as large as 5, getting even larger ($c_2\leq 55$) for  89.90 deg $\leq i\leq 89.99$ deg.
For  $a=7000$ km and 89.0 deg $\leq i\leq 89.9$ deg the period of the node of LARES, which is the same of the  node harmonic perturbation induced by the $\ell=2,m=1$ constituent of the solar $K_1$ tide, not cancelled out by the combination of \rfr{combi}, amounts to several years, as shown by Figure \ref{nodo};
 this fact would induce a serious systematic error because  the $K_1$ tide would act as a superimposed corrupting linear trend over the time spans of some years which would be typically used for the data analysis.

\section{The impact of the inclination errors}
\begin{figure}[htbp]
   \includegraphics[width=17cm,height=13cm]{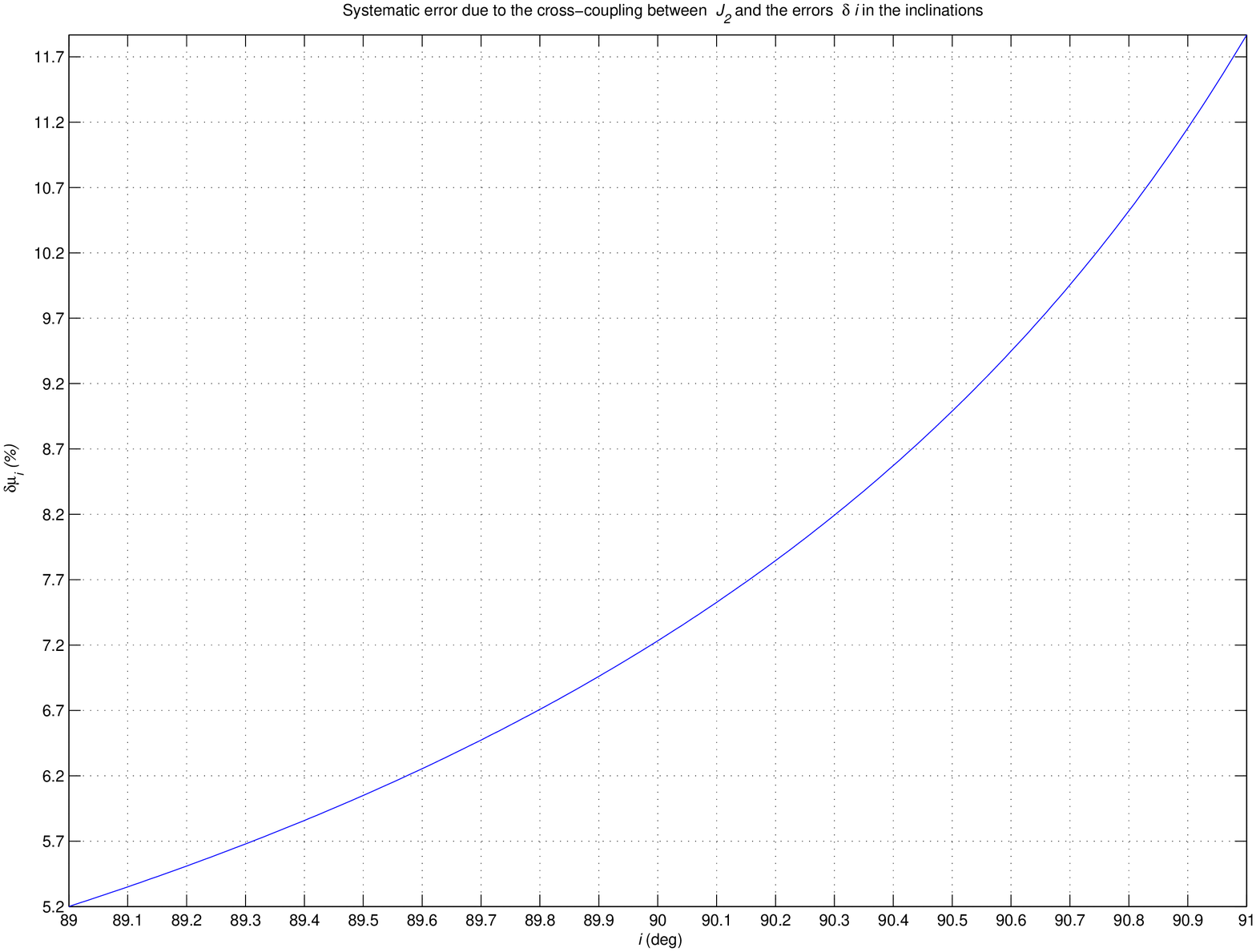}
   \caption{Systematic percent error $\delta\mu_{\rm incli}$ due to the cross-coupling between $J_2$ and the errors $\delta i$ in the inclinations of the satellites entering the combination of \rfr{combi}. We assumed for LARES $a=7600$ km, $e=0.001$, 89 deg $ \leq i\leq 91$ deg.}
   \label{inclazb}
   \end{figure}

Another source of systematic error which may become important for certain orbital configurations is the cross-coupling among $J_2$ and the errors in the inclinations of the satellites
entering the combination  of \rfr{combi}.  Indeed, such a bias $\delta\mu_{\rm incli}$ is not cancelled out by the coefficients of \rfr{cofi}
being equal to
\eqi\delta\mu_{\rm incli}\leq \left(\rp
{\left|-\cf 2{L}J_2\tan i_{\rm L}\delta i_{\rm L}\right| + \left|-
c_1\cf 2{L\ II}J_2\tan i_{\rm L\ II}\delta i_{\rm L\ II }\right|   +\left|-c_2
\cf 2{LR}J_2\tan i_{\rm LR}\delta i_{\rm LR }\right|}
{\left|\dot\Omega^{\rm LAGEOS}_{\rm LT} + c_1\dot\Omega_{\rm LT} ^{\rm LAGEOS\ II}+c_2\dot\Omega_{\rm LT} ^{\rm LARES}\right|}\right)100\lb{errorI}\eqf
If we  assume $\delta i\approx \delta r/a$, with $\delta r\approx 1$ cm for all the satellites, it turns out that \rfr{errorI} yields a further bias which is about $1\%$ for 60 deg $\leq i\leq 80$,
%
 %
 %
 %\begin{figure}[htbp]
 %  \includegraphics[width=17cm,height=13cm]{inclaza.jpg}
 %  \caption{Systematic percent error $\delta\mu_{\rm incli}$ due to the cross-coupling between $J_2$ and the errors $\delta i$ in the inclinations of %he satellites entering the combination of \rfr{combi}. We assumed for LARES $a=7600$ km, $e=0.001$, 60 deg $ \leq i\leq 80$ deg.}
%   \label{inclaza}
%   \end{figure}
 %
 %
 %
 %
but it may become larger for departures of 1 deg from $i=90$ deg, as shown by Figure \ref{inclazb}.

\section{Some considerations on the non-gravitational perturbations and the measurement errors}
It is worthwhile noting that also the impact of the subtle non-gravitational perturbations will be different with respect to the original proposal because LARES will fly in a different and lower orbit and its thermal behavior will  probably be different with respect to  LAGEOS and LAGEOS II. A detailed treatment of this important subject is outside the scopes of this paper; we will only give some insights calling for deeper analyses by more expert researchers in the field. The reduction of the impact of the thermal accelerations, like the Yarkowsky-Schach effects, should have  been reached with two concentric spheres. However, as explained in \cite{Andres}, this solution will increase the floating potential of LARES because of the much higher electrical resistivity and, thus, the perturbative effects produced by the charged particle drag. 
Moreover, drag will increase also because of the lower orbit
of the satellite, both in its neutral and charged components; preliminary
calculation point towards a secular decrease of the inclination of LARES
which maps into a further, uncancelled bias in the node precessions due to
J2 which may be relevant over the typical timescales of the test \cite{LTin}.
Also the Earth's albedo, with its anisotropic components, should have a major effect.

Another point which must be considered is the realistic orbit accuracy obtainable for LARES. Indeed, at a lower orbit the normal points RMS will be probably higher with respect to the present RMS obtained for the two LAGEOS satellites (a few mm), as we presently know for the Stella and Starlette normal points. Of course, such an accuracy is a function of several aspects.

\section{Conclusions}
In this paper we have calculated the systematic error due to the even zonal harmonics of the geopotential, up to degree $\ell=20$, on the measurement of the Lense-Thirring effect to be performed with the existing LAGEOS and LAGEOS II satellites along with the recently approved LARES, which will be launched at the end of 2008 by ASI with a VEGA rocket.

By taking the differences between several Earth's gravity solutions from the dedicated GRACE mission instead of optimistically considering the statistical covariance sigmas of each model separately, as done so far, we have shown that, with the orbital configuration of LARES which should be implemented in such a mission ($a=7600$ km, $60$ deg $\leq i\leq 80$ deg), the claim of reducing the impact of the mismodelling in the even zonals at a $\lesssim 1\%$ level is optimistic because it may be up to ten times larger according to calculations up to degree $\ell=20$. Taking into account the even zonals of degree $\ell >20$ whose action must be considered  because of the relatively low altitude of LARES, may further degrade the total accuracy. Should a  nearly polar, lower-orbit ($a=7000$ km, $i\approx 90$ deg)  is implemented, all the figures would get even worse: in this case, also other orbital perturbations would contribute to corrupt the outcome of the test.
%Moreover, a too low altitude, corresponding to a semimajor axis $a\approx 7000$ km, would further enhance the impact of the static part of the even %zonals which should be taken into account also for degree $\ell\gg 20$.
The orbital configuration of LARES may also have an impact on some non-gravitational perturbations and measurement errors.

Only the originally proposed LARES configuration ($a=12270$ km, $i=70$ deg) would damp the systematic geopotential error to a few percent level.

In conclusion, according to the present-day state-of-the-art in our knowledge of the terrestrial gravity field, the ongoing LARES mission, as it  seems it will be implemented, should not be able to allow for a $\lesssim 1\%$ measurement of frame-dragging.    Maybe LARES will be more useful for studies of geodesy and geophysics.
%
%-----------------------------------------

\end{document}